# Laser Particle Heating Process in a Stand-off Photo-thermal Explosive Detection System

Philip Cassady, PhD
Cassady Engineering Company
pcassady@alum.mit.edu

**Abstract:**

Recent publications have described a method for stand-off optical detection of explosives using resonant infra-red photothermal imaging. This technique uses tuned lasers to selectively heat small particles of explosive lying on a substrate surface. The presence of these heated particles is then detected using thermal infra-red imagery. Although the method has been experimentally demonstrated, no adequate theoretical analysis of the laser heating and subsequent particle cooling has been developed. This paper provides the analytical description of these processes that is necessary to understand and optimize the operational parameters of the explosive detection system. The differential equations for particle and substrate temperatures are derived and solved in the Laplace transform domain. The results are used to describe unexplained cooling phenomena measured during the experiments. A limiting particle temperature is derived as a function of experimental parameters. The effects of radiative and natural convection cooling of the particle and of non-uniform particle temperature are examined and found to be negligible. Calculations using the analytical model are compared with experimental measurements. An analysis of thermal contact heating of the substrate is included in the appendix.

**Introduction:**

An optical method for the detection of trace explosives using photo-thermal infrared spectroscopy has been investigated in several recent papers[1,2,3,4]. The concept relies on laser heating of small particles of explosive that are in contact with a solid substrate. It is necessary to analyze the laser heating process in order to optimize the infrared emission of the small particles by proper choice of laser power and laser pulse length based on the physical properties of the particles and the substrate on which they lie. An analysis is presented in Ref. 1, but it does not model the thermal contact between the particle and the substrate correctly[5]. An improved analysis of the laser heating of a small spherical particle in thermal contact with a substrate surface is presented in this paper. Similar assumptions are made here as those made in Ref. 1: the particle is small enough so that its temperature is uniform throughout, the temperature rise in the particle is not high enough to provide appreciable radiative or convective cooling so radiation and convection heat losses are neglected, and the laser wavelength is chosen to preferentially heat the particle so that laser heating of the substrate material is neglected. The analytical model developed here shows that the characteristic time for particle cooling between laser pulses increases with the laser pulselength. The particle temperature is shown to approach an asymptotic level after a large number of laser pulses are absorbed. Particle temperature non-uniformity and the effects of radiative and natural convection particle cooling are examined and shown to be negligible for typical conditions. Calculations using this analytical model are shown to compare very well with experimental measurements.

**The Analytical Model:**

The situation described in Ref. 1 will be analytically modeled here: a single small (5 micron diameter) spherical particle heated by laser radiation is in thermal contact over a very small area with a thermally conducting substrate that is not heated by the laser. The particle is assumed to be small enough that its temperature is uniform throughout. The necessary physical properties of the particle are: R, the density of the particle material ($kg/m^3$), C the specific heat of the particle material (J/kg/K), V the volume of the spherical particle ($m^3$), h the contact conductance between the particle and the substrate surface ($W/m^2/K$), a the radius of the contact area between the particle and the substrate surface (m), ρ the density of the substrate material ($kg/m^3$), c the specific heat of the substrate material (J/kg/K), and k the thermal diffusivity of the substrate ($m^2$/sec).

Estimates of magnitude are made at several places in this paper in order to simplify the analysis. These estimates are made using the physical properties of a spherical particle of RDX on either a substrate of aluminum (high thermal conductivity) or plastic (low thermal conductivity). The physical properties of these materials are given in Table 1.

| | ***Density*** ***$kg/m^3$*** | ***Specific Heat*** ***J/kg/K*** | ***Particle Volume*** ***$m^3$*** | ***Thermal Diffusivity*** ***$m^2$/sec*** |
|---|---|---|---|---|
| **RDX** | R=1800 | C=1260 | $V=6.55x10^{-17}$ | $k=1.29x10^{-7}$ |
| **Aluminum** | ρ=2700 | c=904 | | $k=8.23x10^{-5}$ |
| **Plastic** | ρ =1190 | c=1465 | | $k=1.2x10^{-7}$ |
| **Polyethylene** | R=950 | C=2200 | $V=5.44x10^{-14}$ | $k=2.29x10^{-7}$ |
| **Copper** | ρ =8960 | c=385 | | $k=1.11x10^{-4}$ |

**Table 1: Material Properties of Particles and Substrates**

The value of h for these estimates was taken from data[6] for various materials on an aluminum substrate:

$$h = 500\ BTU/hr/ft^2/F = 2835\ W/m^2/K \qquad (1)$$

The differential equation for the particle temperature, Tp (K) is given in the first part of Equation 1 in Ref. 1:

$$RCV\frac{\partial T_p}{\partial t} = \dot{q} - h\pi a^2\left[T_p(t) - T_s(0,0,t)\right] \qquad (2)$$

Where the contact heat transfer term has been corrected to include the contact area. It will be assumed in this analysis that the substrate temperature is constant over this very small contact area and is equal to its value at the center of the small contact area.

The center of the contact area is the origin of a cylindrical coordinate system on the surface of the substrate material. The laser heating rate (W) into the particle, q(dot), is the product of the laser intensity falling on the particle (W/m2) times the particle area, times the non-dimensional absorption cross section for the particular laser wavelength and particle material. This absorption cross section is described in Ref. 1 for typical laser wavelengths and particle materials.

The time history of the substrate temperature must be calculated using the appropriate heat diffusion equation in the substrate material. Carslaw and Jaeger[7] provide the solution of this heat diffusion equation for a heat supply at a rate of Q per unit time per unit area (W/m$^2$) over the circle with radius less than or equal to a (m). The temperature at a point whose cylindrical coordinates are r, θ, z the solution is given as:

$$T_s = \frac{aQ}{2K}\int_0^\infty J_0(\lambda r)J_1(\lambda a)\left\{e^{-\lambda z}erfc\left[\frac{z}{2\sqrt{kt}} - \lambda\sqrt{kt}\right] - e^{\lambda z}erfc\left[\frac{z}{2\sqrt{kt}} + \lambda\sqrt{kt}\right]\right\}\frac{d\lambda}{\lambda} \qquad (3)$$

Where the initial temperature at time = 0 is taken as zero degrees. Since the problem is linear, the calculated value of T will be equal to the temperature rise caused by the laser heating regardless of the initial temperature. The substrate conductivity, K, is equal to the product ρck.

The substrate temperature at the origin of the coordinate system, which is the center of the contact area with the particle is:

$$T_s(0,0,0) = T_{cl} = \frac{2Q\sqrt{kt}}{K}\left[\frac{1}{\sqrt{\pi}} - \frac{e^{-\frac{a^2}{4kt}}}{\sqrt{\pi}} + \frac{a}{2\sqrt{kt}}erfc\left(\frac{a}{2\sqrt{kt}}\right)\right] \qquad (4)$$

$T_{cl}$ will now be used to denote the substrate temperature in the contact area.

The contact heat transfer Q depends on the temperature difference between the particle and substrate and the contact area, which makes it a function of time:

$$Q(t) = h\left[T_p(t) - T_{cl}(t)\right] \qquad (5)$$

Since Q is a function of time, its effect on $T_{cl}$ will be cumulative and must be modeled as a convolution over time. This is accomplished by differentiating Q with respect to time and then integrating the individual contributions dQ over time using the dummy variable τ resulting in the integral equation for $T_{cl}$:

$$T_{cl} = \frac{2ha}{K}\int_0^t \frac{\sqrt{k(t-\tau)}}{a}\left\{\frac{1}{\sqrt{\pi}} - \frac{e^{-\frac{a^2}{4k(t-\tau)}}}{\sqrt{\pi}} + \frac{a}{2\sqrt{k(t-\tau)}}\,erfc\left(\frac{a}{2\sqrt{k(t-\tau)}}\right)\right\}\left[\frac{dT_p}{d\tau} - \frac{dT_{cl}}{d\tau}\right]d\tau \tag{6}$$

Using the rules for differentiating an integral equation this can be changed into an equation for $dT_{cl}/dt$:

$$\begin{aligned}\frac{\sqrt{\pi}\rho c}{h}\frac{dT_{cl}}{dt} &= \int_0^t \frac{e^{-\frac{a^2}{4k(t-\tau)}}}{\sqrt{k(t-\tau)}}\frac{dT_{cl}}{d\tau}d\tau - \int_0^t \frac{1}{\sqrt{k(t-\tau)}}\frac{dT_{cl}}{d\tau}d\tau \\ &\quad - \int_0^t \frac{e^{-\frac{a^2}{4k(t-\tau)}}}{\sqrt{k(t-\tau)}}\frac{dT_p}{d\tau}d\tau + \int_0^t \frac{1}{\sqrt{k(t-\tau)}}\frac{dT_p}{d\tau}d\tau\end{aligned} \tag{7}$$

These are all convolution integrals that can be solved with Laplace transforms using the Faltung theorem. Taking the Laplace transforms of this equation gives:

$$\frac{\sqrt{\pi}\rho c}{h}L\left[\frac{dT_{cl}}{dt}\right] = \left\{L\left[\frac{e^{-\frac{a^2}{4kt}}}{\sqrt{kt}}\right] - L\left[\frac{1}{\sqrt{kt}}\right]\right\}\left(L\left[\frac{dT_{cl}}{dt}\right] - L\left[\frac{dT_p}{dt}\right]\right) \tag{8}$$

Taking the Laplace transform of the differential equation for the particle temperature, $T_p$ yields:

$$RCV \cdot L\left[\frac{dT_p}{dt}\right] = L[\dot{q}] - h\pi a^2\left(L[T_p] - L[T_{cl}]\right) \tag{9}$$

Since both the particle temperature and the substrate temperature are assumed to be equal to zero degrees at the beginning of the laser pulse, the Laplace transforms of the differentials with respect to time yield two equations in the two unknowns, L[Tp] and L[Tcl]:

$$L[T_{cl}]\left\{\frac{\rho c}{h}p-\sqrt{\frac{1}{k}}\sqrt{p}e^{-a\sqrt{\frac{p}{k}}}+\sqrt{\frac{1}{k}}\sqrt{p}\right\}+$$

$$L[T_p]\left\{\sqrt{\frac{1}{k}}\sqrt{p}e^{-a\sqrt{\frac{p}{k}}}-\sqrt{\frac{1}{k}}\sqrt{p}\right\}=0 \tag{10}$$

$$L[T_{cl}]\{-h\pi a^2\}+L[T_p]\{pRCV+h\pi a^2\}=L[\dot{q}]$$

These two simultaneous equations can be solved for L[Tp] using Cramer's rule:

$$L[T_p]=\frac{Numerator(p)}{Deno\min ator(p)}$$

$$Numerator(p)=\left\{p\frac{\rho c}{h}+\sqrt{p}\sqrt{\frac{1}{k}}\left(1-e^{-a\sqrt{\frac{p}{k}}}\right)\right\}\{L[\dot{q}]\} \tag{11}$$

$$Deno\min ator(p)=p\left[p\frac{\rho c}{h}RCV+\sqrt{p}\sqrt{\frac{1}{k}}RCV\left(1-e^{-a\sqrt{\frac{p}{k}}}\right)+\pi a^2\rho c\right]$$

**Time Dependence of Particle Temperature**

The time dependence of the particle temperature is determined by the poles of L[Tp] located at the values of p that cause the denominator to vanish. One pole is located at p=0. The remaining poles are located at values of p that satisfy the equation:

$$p\frac{\rho c}{h}RCV+\sqrt{p}\sqrt{\frac{1}{k}}RCV\left(1-e^{-a\sqrt{\frac{p}{k}}}\right)+\pi a^2\rho c=0 \tag{12}$$

This complicated transcendental equation has been approximated by examining two alternative cases. In the first case, the assumption is made that the value of p satisfying Equation 12 is small enough that the exponential can be expanded in terms of the small exponent. Since the radius of the contact area is expected to be much smaller than the square root of the substrate thermal diffusivity, this exponent is expected to be small.

When this is done and the first two terms of the expansion of the exponential are retained, the simplified expression for the denominator has zeroes located at:

$$p = -\frac{\pi a^2 \rho c}{\left(\frac{\rho c}{h} + \frac{a}{k}\right) RCV} \equiv -\frac{1}{\gamma}$$
$$p = 0 \tag{13}$$

When either of these values for p is substituted into the exponential expressions in Equation 12, the assumption to expand the exponentials in terms of their small parameter is seen to be justified.

The alternative case is to assume that p is very large in Equation 12 so that the exponential terms can be neglected. When this is done, the solution to Equation 12 yields two complex values of p. However the magnitudes of these complex roots are again found to be small, not large enough to justify the neglect of the exponential in Equation 12. Therefore the only first alternative case will be developed here.

The zeroes of the denominator determine the temporal behavior of the particle temperature. The zero at p = 0 represents a solution without time dependence, that is constant in time. The zero at $-1/\gamma$ represents an exponentially decaying solution with a characteristic time represented by $\gamma$.

$$\gamma = \frac{RCV}{h\pi a^2}\left(1 + \frac{ah}{\rho ck}\right) \tag{14}$$

This characteristic time is seen to be the characteristic time in Equation (2) for conductive heat transfer through the contact area increased by the heat diffusion process in the substrate material. It is about 60 milliseconds for both the aluminum and the plastic substrates when the radius of the contact surface is 0.5 micron. It gets shorter for larger contact surface and longer for smaller contact surfaces.

**The Full Solution for the Particle Temperature:**

The solution for the time history of the particle temperature requires the inverse Laplace transformation of the equation for L[Tp].

The laser heating term, q(dot), will be modeled as an instantaneous heat source applied at time = 0 and removed at time = $\delta$ representing a laser pulse of temporal length $\delta$. The Laplace transform for such a terminated step function is given as:

$$L[u(t) - u(t-\delta)] = \frac{1 - e^{-p\delta}}{p} \tag{15}$$

Using this to transform the q(dot) term and expanding the exponential in terms of the small parameter in the numerator gives the equation for $L[T_p]$:

$$L[T_p] = \frac{\frac{\dot{q}}{RCV}\left(1 - e^{-p\delta}\right)}{p\left(p + \frac{1}{\gamma}\right)} = \frac{\frac{\dot{q}}{RCV}}{p\left(p + \frac{1}{\gamma}\right)} - \frac{\frac{\dot{q}}{RCV}}{p\left(p + \frac{1}{\gamma}\right)} e^{-p\delta} \qquad (16)$$

The inverse Laplace transform can be either looked up in Reference 8 or calculated using the method of partial fractions yielding the solution for the time dependence of the particle temperature:

$$T_p(t) = \frac{\dot{q}\gamma}{RCV}\left[\left(1 - e^{-t/\gamma}\right) - U(t-\delta)\left(1 - e^{-(t-\delta)/\gamma}\right)\right] \qquad (17)$$

Where U(t-δ) is the unit step function that begins when t= δ (the end of the laser pulse).

As the radius of the contact surface gets smaller, the characteristic time, γ, gets larger and the particle temperature history approaches the linear growth of temperature with time that is imposed by the laser heating alone without thermal contact with the substrate.

$$T_{p,NoContact}(t) = \frac{\dot{q}}{RCV} t \qquad (18)$$

A particle time history is shown below for the 5 micron diameter RDX particle on a plastic substrate. The material properties are given earlier in Table 1. The radius of the contact surface is assumed to be 1/2 micron. A 100 $mW/cm^2$ laser fully absorbed over the particle cross section heats the particle for 10 msec.

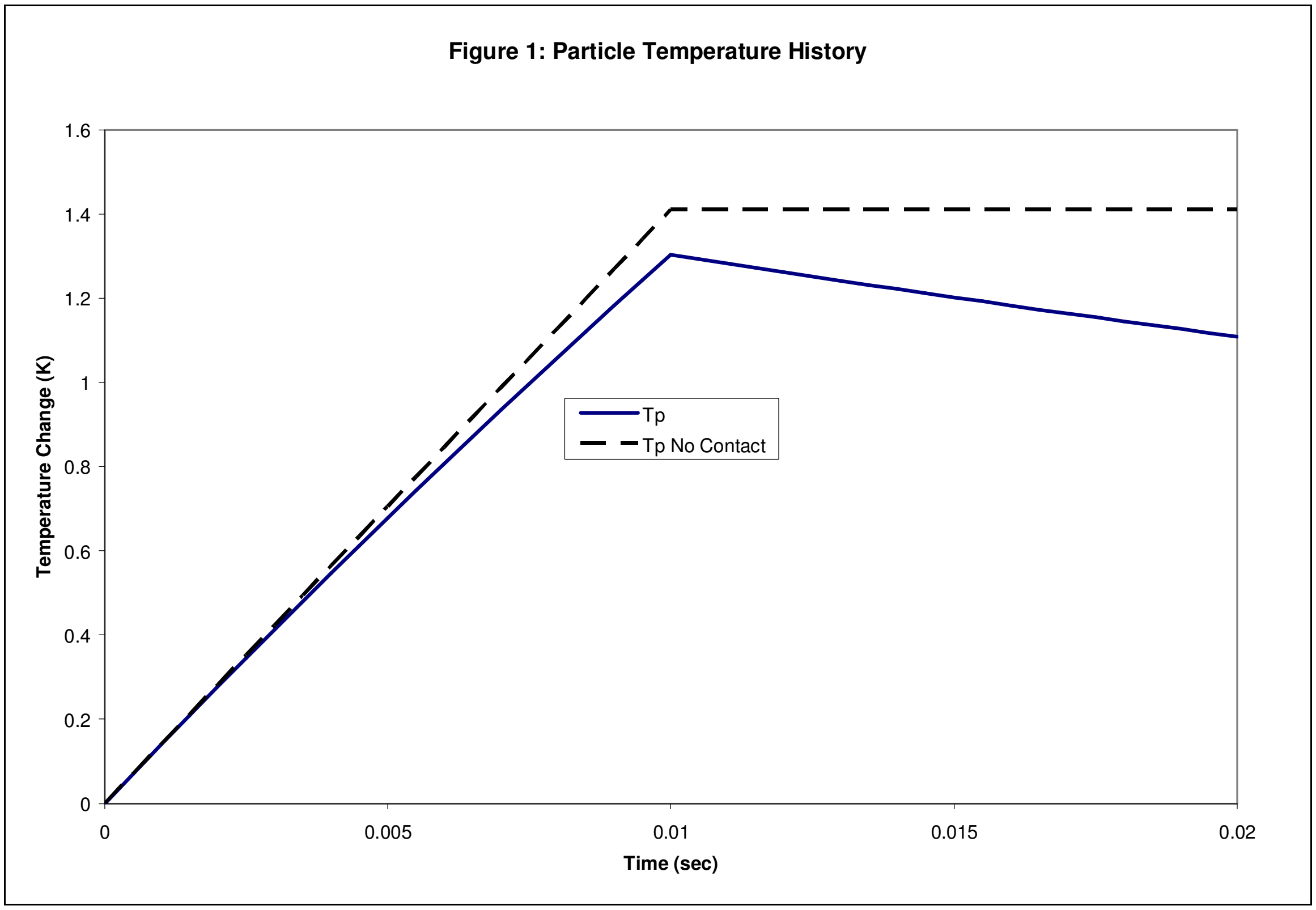

**Figure 1: Particle Temperature History with 10 msec Laser Pulse and Contact Conduction to Substrate**

Examination of this figure shows that the heating process occurs over a much shorter time scale than the cooling process. Under these conditions, it takes about 5 msec to raise the temperature 0.5 degK, but after the laser turns off, it takes more than 10 msec for the temperature of the particle to drop 0.5 degK from its peak.

The experimental observation that the time required for the particle to lose 10% of its peak temperature grew larger as the laser pulse length was increased was unexplained in Ref. 2. This effect can be explained using the results of the present analysis. The time required for the particle temperature to drop 10% below its peak temperature can be expressed using Equation (17) as:

$$t(cool10\%) = \frac{10}{9}\gamma \cdot e^{\delta/\gamma} \tag{19}$$

For the conditions described above this 10% cooling time is shown to grow as a function of the laser pulselength in Figure 2.

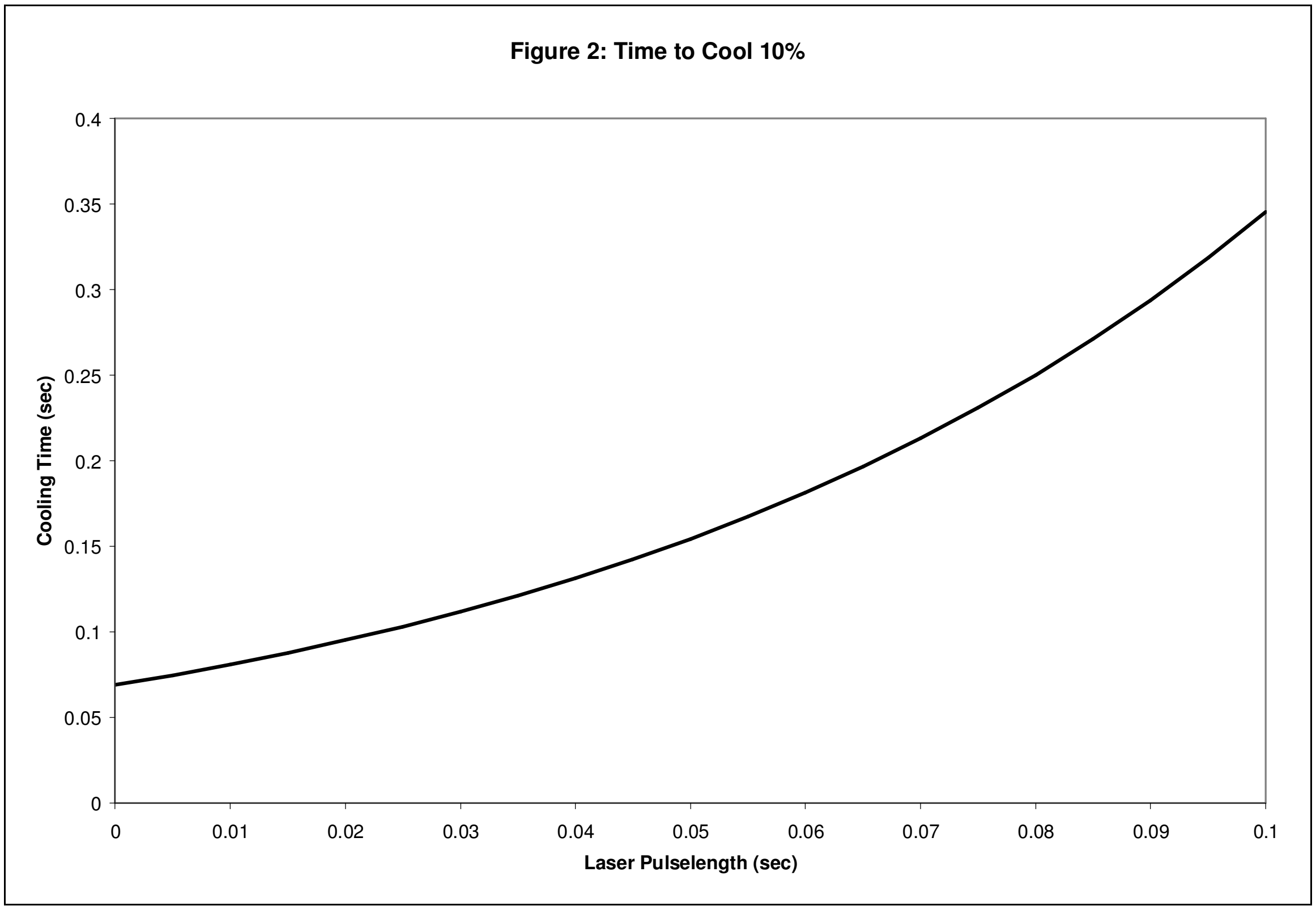

**Figure 2: Time Required for Particle to Cool 10% by Contact Conduction to Substrate**

## Particle Temperature after N Laser Pulses

As each laser pulse is absorbed the temperature of the particle continues to rise until a steady state is reached in which the amount of energy absorbed by the particle in each subsequent laser pulse is equal to the amount of energy transferred to the substrate during the pulse and the cooling off period that follows the pulse. Assuming that the duty cycle of the laser is 50% the peak particle temperature at the end of the Nth pulse can be calculated as:

$$Tpp(N) = \frac{\dot{q}\gamma}{RCV} \sum_{n=0}^{2N-1} (-1)^n e^{-n\delta/\gamma} \qquad (20)$$

For the conditions given above, this peak particle temperature after N pulses is given in Figure 3:

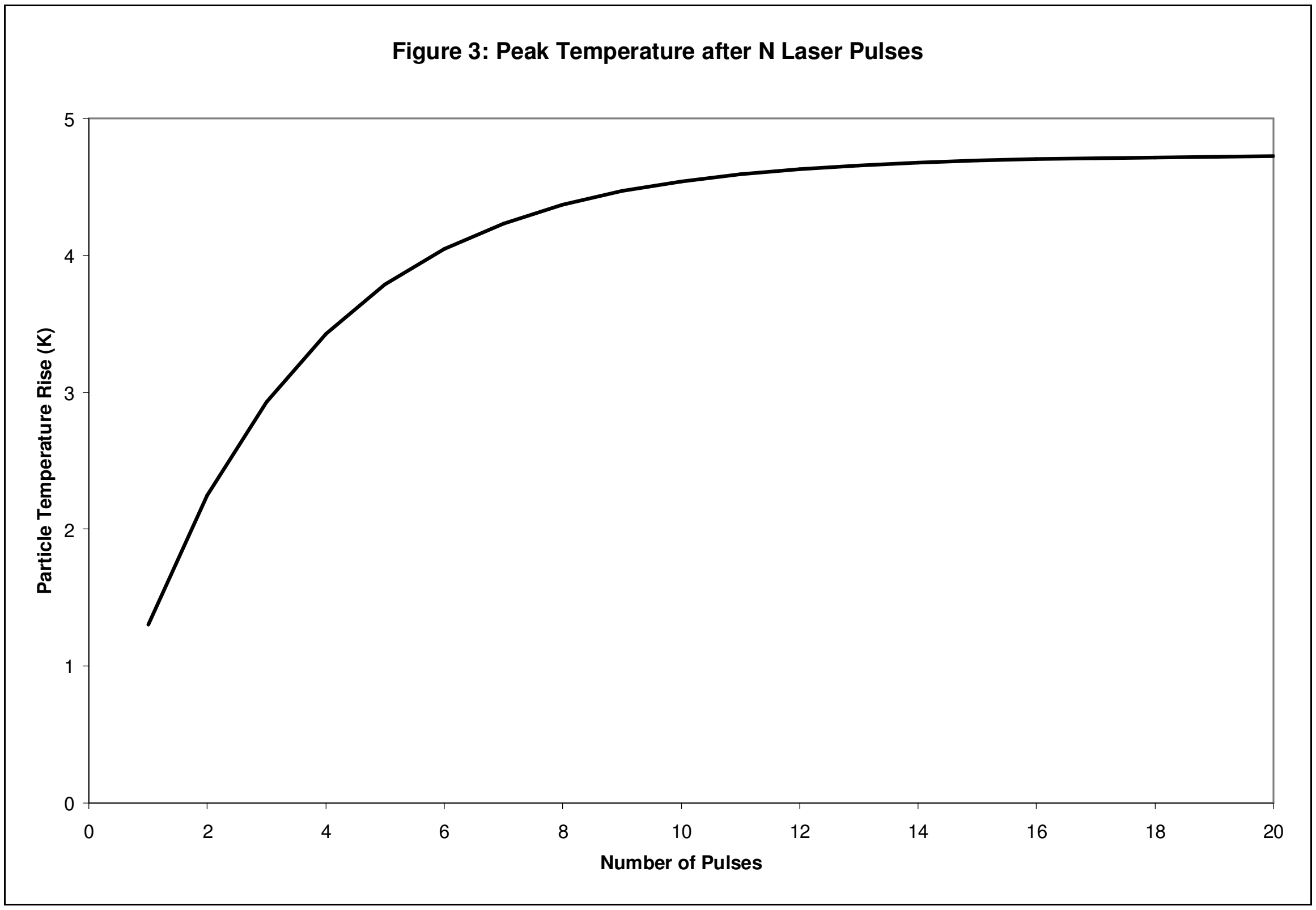

**Figure 3: Peak Particle Temperature after N Laser Pulses**

After approximately 20 laser pulses, the rise in particle temperature is seen to approach a limiting value of approximately 4.7 K above ambient. In the limit of a very large number of pulses, Equation (20) can be summed as a binomial series:

$$Tpp\lim = \frac{\dot{q}\gamma}{RCV}\left(\frac{1}{1+e^{-\gamma/\delta}}\right) \tag{21}$$

**Effect of Radiative and Natural Convection Cooling:**

The radiative cooling of a particle is described by the differential equation:

$$RCV\frac{dT_{prad}}{dt} = -\varepsilon\sigma\left(4\pi r_{ple}^2\right)\left(T_{prad}{}^4 - T_{amb}{}^4\right) \tag{22}$$

Where $r_{ple}$ is the particle radius and $T_{amb}$ is the ambient temperature of the surroundings that are exchanging radiative energy with the particle, $T_{prad} = T_{amb} + T_p$, ε is the particle emissivity (taken as 1 here) and σ is the Stefan-Boltzmann constant. Since $T_p$ is expected to be much smaller than $T_{amb}$, the linearized version of the radiative cooling equation is:

$$RCV\frac{dT_p}{dt} = -\frac{\varepsilon\sigma 16\pi r_{ple}^{\ \ 2}}{T_{amb}} T_p \qquad (23)$$

Particle cooling by natural convection is described by the differential equation:

$$RCV\frac{dT_{conv}}{dt} = -\frac{k_{air}Nu}{2r_{ple}}\left(4\pi r_{ple}^2\right)\left(T_{conv} - T_{amb}\right)$$
$$Nu = 0.49(Gr \cdot \Pr)^{1/4}$$
$$Gr = \frac{g(T_{conv} - T_{amb})(2r_{ple})^3}{T_{amb}\nu_{air}^2} \qquad (24)$$
$$\Pr = \frac{\mu_{air}c_{pair}}{k_{air}}$$

The Nusselt number for low speed natural convection over a sphere was taken from page 6-15 of Reference 5. $T_{conv}$ is equal to the ambient temperature, $T_{amb}$, plus the elevation of the particle temperature above ambient, $T_{p.}$

In order to linearize the equation for natural convection particle cooling, the particle temperature appearing in the Grashof number will be assumed to be equal to the particle temperature that would have occurred without any radiative or natural convective cooling. The effect of the Grashof number is small because of the small exponent in the Nusselt number relation. This assumption will be later justified by the fact that radiative and natural convective cooling have very little effect on the particle temperature.

The linearized equation for natural convective cooling of the particle can be written as:

$$RCV\frac{dT_p}{dt} = -k_{air}Nu \cdot 2\pi r_{ple}\frac{A_{cov}}{A}T_p \qquad (25)$$

The term $A_{cov}/A$ is the fraction of effective particle area included to allow shadowing of the particle by the nearby substrate surface. Its value was taken as 0.5 in the calculations below.

The radiative cooling and natural convection cooling effects can be combined into a single additional term that will appear on the right hand side of the particle temperature differential equation, Equation (2). This equation will now appear as:

$$RCV\frac{\partial T_p}{\partial t} = \dot{q} - h(\pi a^2)[T_p(t) - T_s(0,0,t)] - Const_s T_p$$

$$Const_s = \frac{\varepsilon\sigma 16\pi r_{ple}^2}{T_{amb}} + k_{air}2\pi r_{ple}\frac{A_{\mathrm{cov}}}{A}Nu \qquad (26)$$

The only effect of substitution of this new differential equation for particle temperature into the Laplace transform process described above is to change the characteristic time in the problem from the value given in Equation (14) to:

$$\lambda = \frac{RCV\left(1+\frac{ah}{\rho ck}\right)}{h\pi a^2 + Const_s\left(1+\frac{ah}{\rho ck}\right)} \qquad (27)$$

The inclusion of radiative and natural convective cooling has the effect of reducing the characteristic time.

Substituting λ for the characteristic time in Equation (17) shows that asymptotic long time particle temperature rise with radiative and natural convection cooling is 98% of the asymptotic long time particle temperature rise without radiative and natural convection cooling for the conditions described above. It can be concluded that radiative and natural convection cooling have little effect on particle temperature rise for this problem.

**Uniformity of Particle Heating:**

The particle temperature was assumed to be uniform throughout this analysis. The Fourier number is the ratio of the particle heating time to the thermal accommodation time in the particle. For the 5 micron diameter RDX particle heated for 10 milliseconds by the laser, the Fourier number is:

$$Fo = \frac{k_{ple}\delta}{r_{ple}^2} = 51.6 \qquad (28)$$

So that thermal conduction in the particle has adequate time to provide a uniform particle temperature during the laser pulse heating.

**Comparison with Experimental Data:**

The Naval Research Lab has performed experiments[9] using 20 to 27 micron diameter polyethylene beads on an optically polished copper substrate. The beads were doped with graphite to absorb 658 nm laser radiation at an intensity of 7.6 kW/m$^2$. The particle temperatures were recorded with a single channel mercury-cadmium-telluride detector with DC coupled pre-amplifier.

The material constants for a 23.5 micron diameter polyethylene particle and a copper substrate required to calculate the particle temperature rise using Equation (17), and the characteristic heating time using Equation (14) are given in Table 1.

The experimental data provides the time history of photothermal signal in volts versus the time. The voltage recorded by the detector is proportional to the particle temperature rise. The experimental data were curve fitted using Equation (17) to derive an experimental value for γ, which was found to be 6.99 msec. The quality of the curve fit is shown in Figure 4.

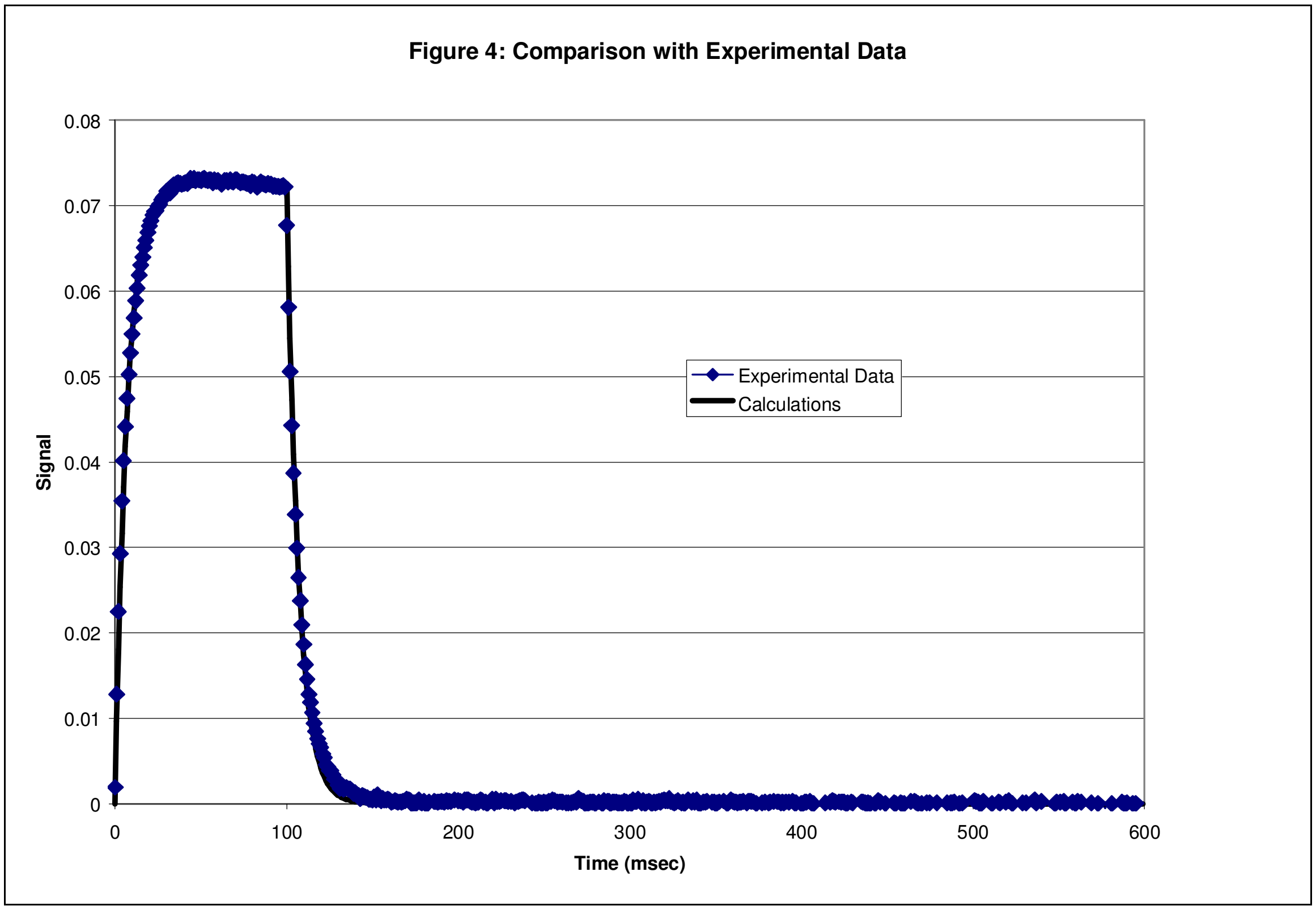

**Figure 4: Comparison with Experimental Data**

Equation (14) can be solved for h to yield:

$$h = \frac{RCV}{\left(\gamma\pi a - \frac{RCV}{\rho c k}\right)a} \tag{29}$$

Which has a value of 7985 $W/m^2/K$ for a 9 micron contact radius using the material parameters given in Table 1. Since the thermal conductivity of copper is twice as large as that of aluminum, it may be expected that this value would be larger than the value assumed for aluminum in Equation 1.

**Conclusions:**

An analytical solution for the thermal time history of a small particle in thermal contact with a conductive substrate has been derived from basic heat transfer theory. The assumptions underlying the model have been examined and verified. The predictions of this analytical model compare well with experimental results. This analytical model can now be used to optimize the design and operation of a photothermal explosive detection system.

**Appendix A: Calculation of Substrate Temperature:**

Experimental measurements indicate that the temperature of the substrate may affect the particle temperature measurement, particularly for a substrate with low thermal conductivity. The substrate surface temperature can be calculated using a technique similar to the particle temperature calculation.

The temperature distribution on the substrate surface can be written using Equation 3 as:

$$T_s(r,0,t) = \frac{aQ}{\rho ck}\int_0^\infty J_0(\lambda r)J_1(\lambda a)erf(\lambda\sqrt{kt})\frac{d\lambda}{\lambda} \quad \text{(A1)}$$

Where Q is the heat input (W/m$^2$) to the surface through the area with radius a from the heated particle.

$$Q(t) = h\cdot[T_p(t) - T_{cl}(t)] \quad \text{(A2)}$$

Since Q is a function of time, the equation for $T_s$ must be written for each elemental change in heat input and the resulting convolution integral integrated over the elemental changes in heat input:

$$T_s(r,0,t) = \int_0^t \frac{a}{\rho c k} \frac{dQ}{d\tau} \int_0^\infty J_0(\lambda r) J_1(\lambda a) erf(\lambda \sqrt{k(t-\tau)}) \frac{d\lambda}{\lambda} d\tau \qquad \text{(A3)}$$

The time dependence of the particle temperature was calculated earlier as:

$$T_p(t) = \frac{\gamma \dot{q}}{RCV}(1 - e^{-t/\gamma}) \qquad \text{(A4)}$$

Using a similar technique, the time dependence of the substrate centerline temperature can also be calculated:

$$T_{cl}(t) = \frac{\dot{q}}{\pi a k \rho c}\left(1 - e^{-t/\gamma}\right) \qquad \text{(A5)}$$

So that the heat input and its derivative are:

$$\begin{aligned} Q(t) &= \frac{\dot{q}}{\pi a^2}\left(1 - e^{-t/\gamma}\right) \\ \frac{dQ}{dt} &= \frac{\dot{q}}{\pi a^2 \gamma} e^{-t/\gamma} \end{aligned} \qquad \text{(A6)}$$

Substituting A6 into A4, introducing y = a λ, and rearranging the integrals yields:

$$T_s(r,0,t) = \frac{\dot{q}}{\rho c k \pi a \gamma} \int_0^\infty J_0\left(y \frac{r}{a}\right) J_1(y) \left[\int_0^t e^{-\tau/\gamma} erf\left(\frac{y}{a}\sqrt{k(t-\tau)}\right) d\tau\right] \frac{dy}{y} \qquad \text{(A7)}$$

The time integral can be integrated by parts to yield:

$$T_s(r,0,t)=\frac{\dot{q}}{\rho ck\pi a}\left\{\begin{array}{l}\displaystyle\int_0^\infty J_0\left(y\frac{r}{a}\right)J_1(y)\,erf\left(\frac{y}{a}\sqrt{kt}\right)\frac{dy}{y}-\\[2ex]\displaystyle\frac{\sqrt{k}e^{-t/\gamma}}{a}\int_0^\infty J_0\left(y\frac{r}{a}\right)J_1(y)\frac{erf\left(\sqrt{\left(\frac{y^2k}{a^2}-\frac{1}{\gamma}\right)t}\right)}{\sqrt{\left(\frac{y^2k}{a^2}-\frac{1}{\gamma}\right)}}dy\end{array}\right\} \tag{A8}$$

It should be noted that the argument of the error function in the second integral is imaginary for values of y between 0 and a/sqrt(k $\gamma$). The radical in the denominator is also imaginary in this range of y.

It is of interest to calculate the average substrate temperature within a circular area of radius R. This average temperature can be defined as:

$$\overline{T}_s(R,0,t)=\int_0^R\frac{2\pi rT_s(r,0,t)}{\pi R^2}dr \tag{A9}$$

Since the r variable only appears in the $J_0$ Bessel function, this integration can be easily carried out yielding:

$$\overline{T}_s(R,0,t)=\frac{2\dot{q}}{\rho ck\pi R}\left\{\begin{array}{l}\displaystyle\int_0^\infty J_1\left(y\frac{R}{a}\right)J_1(y)\,erf\left(\frac{y}{a}\sqrt{kt}\right)\frac{dy}{y^2}-\\[2ex]\displaystyle\frac{\sqrt{k}e^{-t/\gamma}}{a}\int_0^\infty J_1\left(y\frac{R}{a}\right)J_1(y)\frac{erf\left(\sqrt{\left(\frac{y^2k}{a^2}-\frac{1}{\gamma}\right)t}\right)}{\sqrt{\left(\frac{y^2k}{a^2}-\frac{1}{\gamma}\right)}}\frac{dy}{y}\end{array}\right\} \tag{A10}$$

Calculations were done for the small (23.5 micron average diameter) polyethylene beads on a polyethylene substrate. The radius of the contact area and the contact heat transfer coefficient were taken from the experimental data and analysis of the same polyethylene beads on copper substrate (9 microns and 7895 $W/m^2/K$). The laser heating was 7600 $W/m^2$ for 100 msec.

The figure shows the calculated particle temperature, $T_p$, ten times the calculated centerline substrate temperature, $T_{cl}$, and 20 times the calculated average substrate temperature within a radius or 45 microns, equal to 5 times the contact radius.

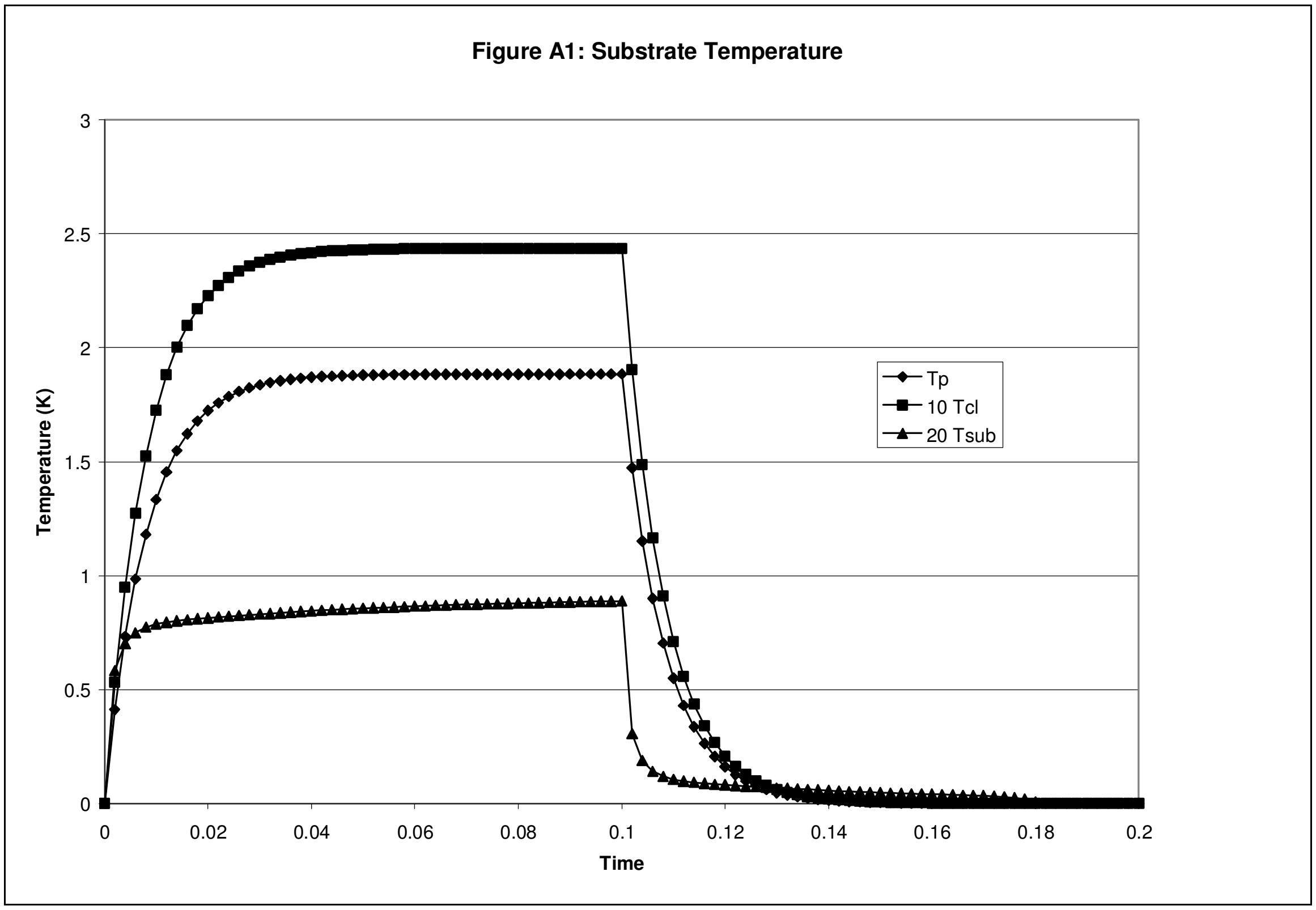

**Figure A1: Substrate Temperature**

The temperatures on the substrate are seen to follow the time history of the particle temperature with the centerline temperature reaching a level of approximately 15 percent of the particle temperature and the average substrate temperature within a 45 micron radius reaching a level of approximately 2 percent of the particle temperature during the laser pulse.